\documentclass[11pt]{article}
\pdfoutput=1 
             
\def\be{\begin{equation}}
\def\ee{\end{equation}}
\def\bea{\begin{eqnarray}}
\def\eea{\end{eqnarray}}

\usepackage{caption}
\usepackage{lineno}
\usepackage{amsmath}
\usepackage{graphicx}
\usepackage{mathtools}
\usepackage{siunitx}
\usepackage{url}
\usepackage{authblk}

\title{Tomographic Muon Imaging of the Great Pyramid of Giza}

\author[1]{Alan D. Bross} 
\author[2]{E.C. Dukes} 
\author[2]{Ralf Ehrlich}
\author[2]{Eric Fernandez}
\author[3]{Sophie Dukes}
\author[4]{Mohamed Gobashy}
\author[5]{Ishbel Jamieson}
\author[6]{Patrick J La Rivi\`{e}re}
\author[6]{Mira Liu}
\author[7]{Gregory Marouard} 
\author[7]{Nadine Moeller}
\author[1]{Anna Pla-Dalmau }
\author[1]{Paul Rubinov} 
\author[8]{Omar Shohoud} 
\author[6]{Phillip Vargas} 
\author[8]{Tabitha Welch}
\affil[1]{Fermi National Accelerator Laboratory, P.O. Box 500, Batavia, IL, USA}
\affil[2]{Physics Department, University of Virginia, Charlottesville, VA, USA}
\affil[3]{Virginia Tech University, Blacksburg, WV, USA}
\affil[4]{Geophysics Department, Faculty of Science, Cairo University, Cairo, Egypt}
\affil[5]{Department of Physics, University of Oxford, Oxford, UK}
\affil[6]{Department of Radiology, University of Chicago, Chicago, IL USA}
\affil[7]{Department of Near Eastern Languages \& Civilizations Yale University, New Haven, CT USA}
\affil[8]{Department of Physics, University of Chicago, Chicago, IL USA}

\begin{document}

\maketitle

\begin{abstract}
The pyramids of the Giza plateau have fascinated visitors since ancient times and are the last of the Seven Wonders of the ancient world still standing.  It has been half a century  since Luiz Alvarez and his team used cosmic-ray muon imaging to look for hidden chambers in Khafre’s Pyramid. Advances in instrumentation for High-Energy Physics (HEP)  allowed a new survey, ScanPyramids, to make important new discoveries at the Great Pyramid (Khufu) utilizing the same basic technique that the Alvarez team used, but now with modern instrumentation.  The Exploring the Great Pyramid Mission plans to field a very-large muon telescope system that will be transformational with respect to the field of cosmic-ray muon imaging.  We plan to field a telescope system that has upwards of 100 times the sensitivity of the equipment that has recently been used at the Great Pyramid, will image muons from nearly all angles and will, for the first time, produce a true tomographic image of such a large structure.
\end{abstract}

\section{Introduction}
\label{sec:Intro}
It has been nearly half a century  since Luiz Alvarez~\cite{Alvarez1z} and his team used cosmic-ray muon imaging to look for hidden chambers in Khafre’s Pyramid. Advances in instrumentation for High-Energy Physics (HEP) have now allowed the ScanPyramids team~\cite{Morishima:2017ghw} to make important new discoveries at the Great Pyramid (Khufu), utilizing the same basic technique that the Alvarez team used, but now with modern instrumentation.  Although there are other modern methodologies that can be used to ``interrogate'' the internal structure of the Great Pyramid, the ScanPyramids team chose to use the same technique employed by Alvarez's team, because cosmic-ray muon imaging presents a very powerful tool that can see deep into the structure.  As in Alvarez's case, they set out looking for ``hidden'' chambers or voids.

Egypt's two largest pyramids, the Great Pyramid of Khufu and the second Giza pyramid of Khafre, look much the same, but the known internal structure of these two pyramids differ remarkably. The  internal spaces of the Great Pyramid are far more complex, most notably demonstrated by its spectacular Grand Gallery, an 8-meter high passageway with corbelled masonry walls.  In addition, micro-gravity measurements made in the 1980s indicate interesting variations in the density of the Great Pyramid high above grade~\cite{BuiHD}.  The Exploring the Great Pyramid (EGP)~\footnote{Approved by the Egyptian Ministry of Antiquities in 2018.} Mission presents the potential to obtain entirely new results on the internal structure of the Great Pyramid by having the precision to measure differences in density.  The precise construction techniques for the Great Pyramid are still relatively unknown with many untested concepts having been hypothesized. Data from the EGP Mission will test many of these hypotheses.

\begin{figure}[h]
\begin{center}
\includegraphics[width=0.75\textwidth]{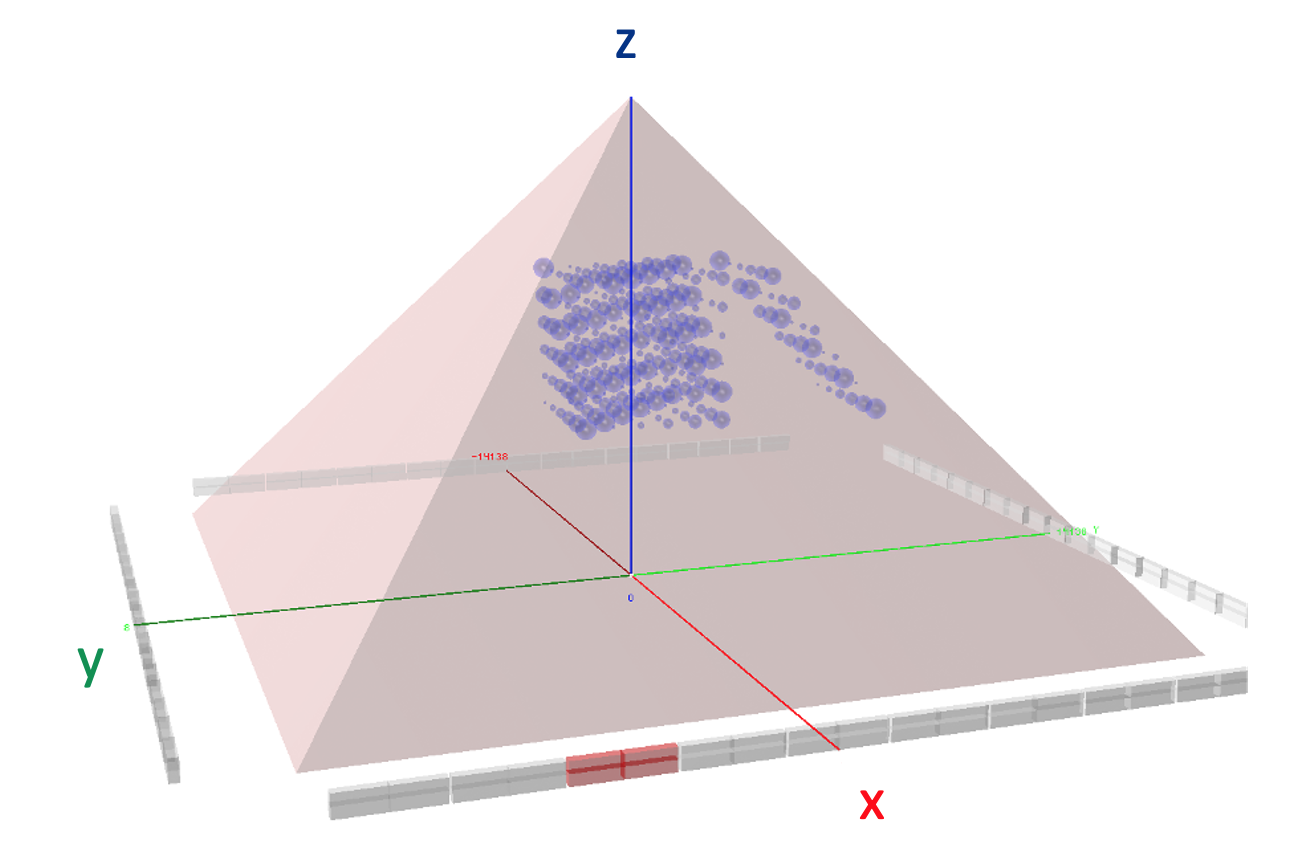}
\caption{GEANT4 model.  The model consists of a solid pyramid with voids with diameters between 1 and 6~m}%
\label{fig:GModel}
\end{center}
\end{figure}

The key elements and strengths of the EGP mission are enumerated below.
\begin{enumerate}
\item Produce a detailed analysis of the entire internal structure which does not just differentiate between stone and air, but can measure variations in density.
\item Answer questions regarding construction techniques by being able to see relatively small structural discontinuities.
\item The large size of the telescope system yields not only the increased resolution, but enables fast collection of the data, which minimizes the required viewing time at the site.  A 2-year view time is anticipated.
\item The telescope is very modular in nature.  This makes it very easy to reconfigure and deploy at another site for future studies.  
\item From a technical perspective, the system being proposed uses technology that has been largely engineered and tested and presents a low risk approach.
\end{enumerate}
Since the detectors that are proposed are very large, they cannot be placed inside the pyramid, therefore our approach is to put them outside and move them along the base.  In this way, we can collect muons from all angles in order to build up the required data set of on $\mathcal{O}~{10^{11}}$ good muons. We are currently in the first phase of EGP where we are performing comprehensive Monte Carlo studies of the proposed system and are developing the needed tomographic reconstruction algorithms. Figure~\ref{fig:GModel} shows our GEANT4~\cite{Ivanchenko:2003xp} training model and one of the telescopes in red.  The objects in grey indicate subsequent positions of the telescope as it is moved along the base. The details of our simulation work are presented in section~\ref{sec:Sim}.
\section{Tomographic imaging with muons}
\label{sec:Tomo}
\subsubsection*{Theoretical considerations}

To perform tomographic reconstruction of the pyramid interior, we
are seeking to leverage the decades of algorithm development for computed
tomography in the medical imaging community. While medical CT scanners
have employed a variety of acquisition geometries (parallel beam,
fanbeam, conebeam), they all rely on the fact that a CT scanner yields
samples of the \emph{line integrals} through the patient's linear
attenuation map along a large set of lines spanning at least 180 degrees
for every slice of the patient. 

In the case of the pyramid, we seek to reconstruct a volumetric image
of the density of the material, $\rho(x,y,z)$, and so need to convert
the measured flux of muons along a given line of sight into a line
integral through this density map. This line integral is referred
to as the \emph{opacity} in the muon tomography literature

\begin{equation} \label{opacity}
\varrho\left(L\right)=\int_{L}\rho(x,y,z)dl.
\end{equation}
Here $L$ denotes an arbitrary line through the pyramid and $dl$
is a differential length along that line. Note that the opacity has
units of $g/cm^{2}$. 

\begin{figure}[h]
\begin{centering}
\includegraphics[width=0.49\textwidth]{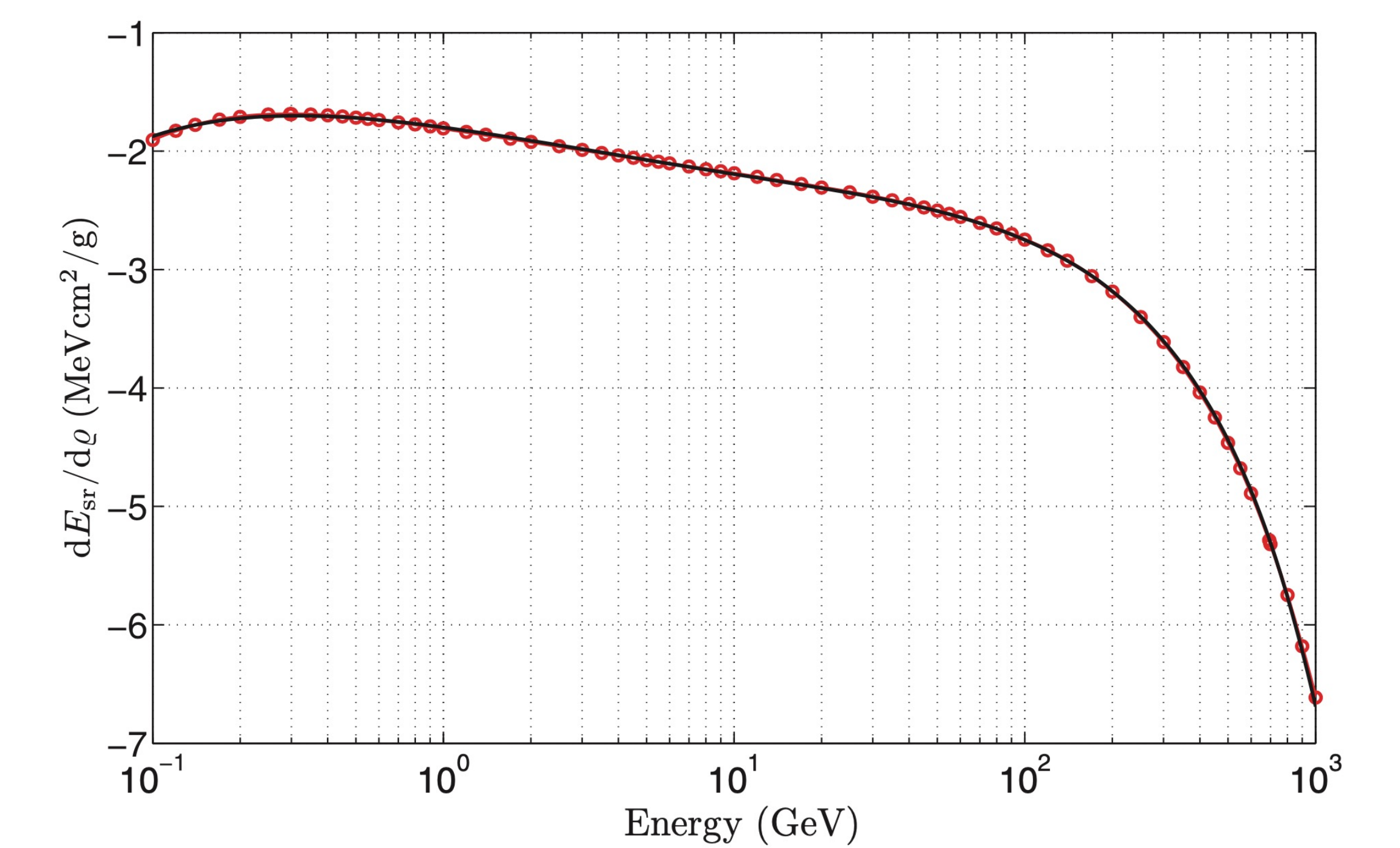}
\includegraphics[width=0.49\textwidth]{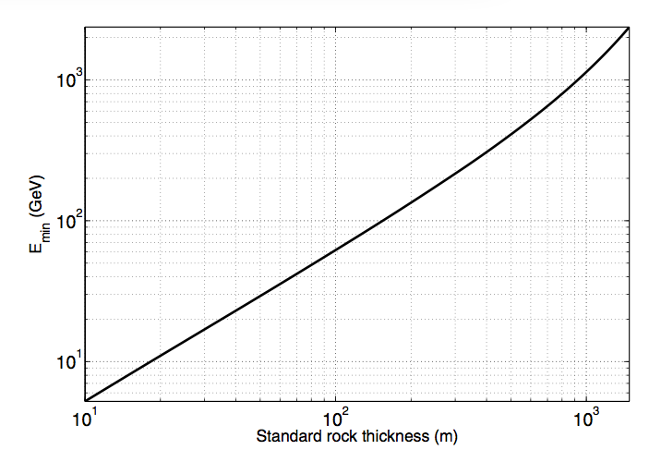}
\par\end{centering}
\caption{\label{fig:Mass-stopping-power}Left: Mass stopping power vs. energy in standard
rock for muons. From Lesparre et al, Geophysical muon imaging: feasibility
and limits.~\cite{Lesparre:2010zz}, Right: Relationship between $E_{min}$
and standard rock thickness (which maps to opacity by multiplying
by density of standard rock).~\cite{Lesparre:2010zz}}
\end{figure}

Unlike x-rays, which interact with a material with some probability
per unit length, charged particles like muons lose energy continuously
as they travel. The energy loss per unit opacity is called the \emph{mass
stopping power} and we will denote it as
\begin{equation} \label{stoppingpower}
T(E)=\frac{dE}{d\varrho}.
\end{equation}
It is tabulated for different materials and a plot of mass stopping power
for standard rock as a function of energy is shown in Figure~\ref{fig:Mass-stopping-power} (Left).

A key point is that for a given incident muon energy, there is a maximum
opacity the muon will cross before running out of energy and being
absorbed. Conversely, for a given opacity, there is a minimum energy
needed for a muon to pass through without being absorbed. These are
related through the following integral equation: 

\begin{equation} \label{opacity-EMin}
\varrho=\int_{0}^{E_{min}}\frac{1}{T(E)}dE.
\end{equation}
This equation implicitly defines a mapping between $\varrho$ and
$E_{min}$ that can be tabulated. A typical such plot is shown in
Figure~\ref{fig:Mass-stopping-power} (Right), where $E_{min}$ is plotted vs. rock thickness, which is related to opacity through the density of standard rock.
%

For a given line of sight, our detection system will count all the
muons whose initial energy is above this $E_{min}$. The spectrum
of incident muon energies along a given line is something that can
be obtained from a model (such as Gaisser~\cite{Gaisser:2008zz}) and with in situ calibration. We denote this by $\Phi(E_{0})$ and it has units of muons ${cm}^{-2} {Sr}^{-1} {s}^{-1} {GeV}^{-1}$
The number of detected muons along a line of opacity $\varrho$
will then be given by

\begin{equation}\label{eq:num_muons}
n\left(\varrho\right)=A_{b}T\Omega_{b}\int_{E_{min}\left(\varrho\right)}^{\infty}\Phi\left(E_{0}\right)dE_{0}
\end{equation}
where $A_b$ is the area of the detection bin, $T$ the measurement time, and $\Omega_b$ the solid angle captured by the detection bin. 
So given the measurement $n(\varrho)$ and the spectral model $\Phi\left(E_{0}\right)$,
the mathematical challenge is to find the $E_{min}\left(\varrho\right)$
that yields the above equality. Then we can use the tabulated mapping
to invert $E_{min}\left(\varrho\right)$ to solve for $\varrho$.
If we can do so for a sufficient number of lines through the pyramid,
we can use tomographic reconstruction strategies from medical imaging
to solve for the 3D density map, which would reveal voids (areas of
density zero) as well as areas of mortar or rubble between intact
stones (areas of reduced density).

\subsubsection*{Practical considerations}

\begin{figure}[t]
\begin{centering}
\includegraphics[width=0.4\columnwidth]{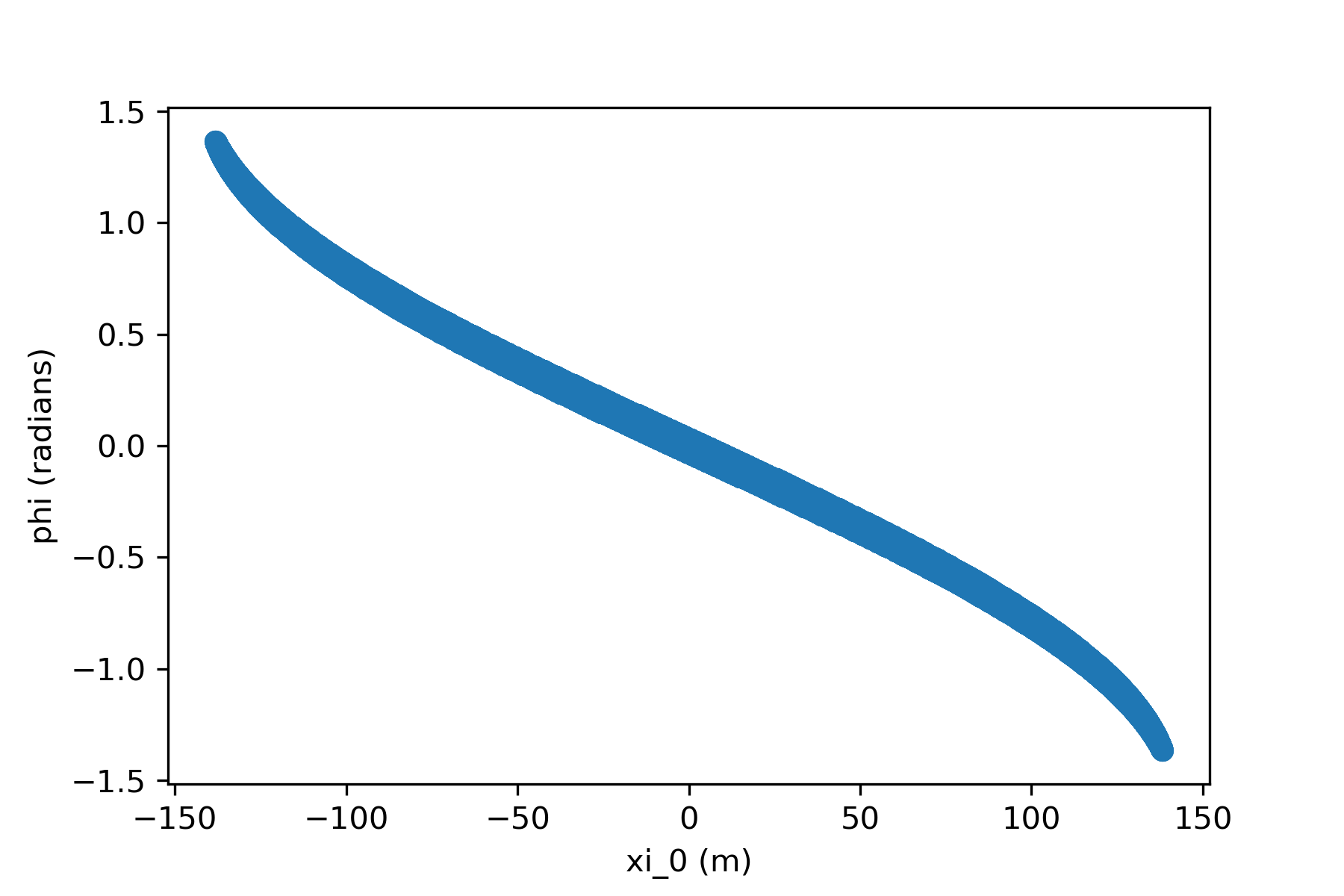}
\includegraphics[width=0.4\columnwidth]{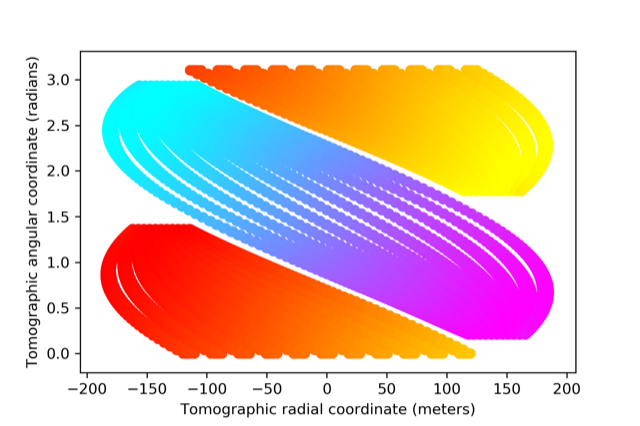}
\par\end{centering}
\caption{\label{fig:Tomographic-sampling-for}Tomographic sampling for (Left)
a single position of the detector and (Right) translation of the detector
to 11 positions along each of two faces of the pyramid. The filling
of the tomographic space on the right would allow for standard tomographic
reconstructions algorithms to be used, although more modern algorithms
may be robust to varying degrees of under-sampling. }
\end{figure}

Given the relatively low flux of cosmic ray muons, we need detectors
with large acceptance ($cm^{2}-steradian).$ We calculated these for
the proposed detectors and found them to be significantly higher than
for existing detectors. 
%
%
Finally, for tomographic reconstruction, it is necessary that the
space of sampled line integrals spans the whole breadth of the pyramid,
as well as at least 180 degrees of rotation in tomography space. The single-slice
sampling of tomography space engendered by a single position of the
detector is shown in Figure~\ref{fig:Tomographic-sampling-for} (Left).
For well-conditioned tomographic reconstruction, this space should
be filled and it is obviously grossly under filled. However, if the
detector is translated along two sides of the pyramid, we obtain the
essentially complete sampling shown in Figure~\ref{fig:Tomographic-sampling-for} (Right).
This analysis shows that significant translation
along two sides of the pyramid will be needed for standard tomographic
algorithms to work.
\section{EGP technical approach}
\label{sec:Tech}
\begin{figure}[h]
\begin{center}
\includegraphics[width=0.40\textwidth]{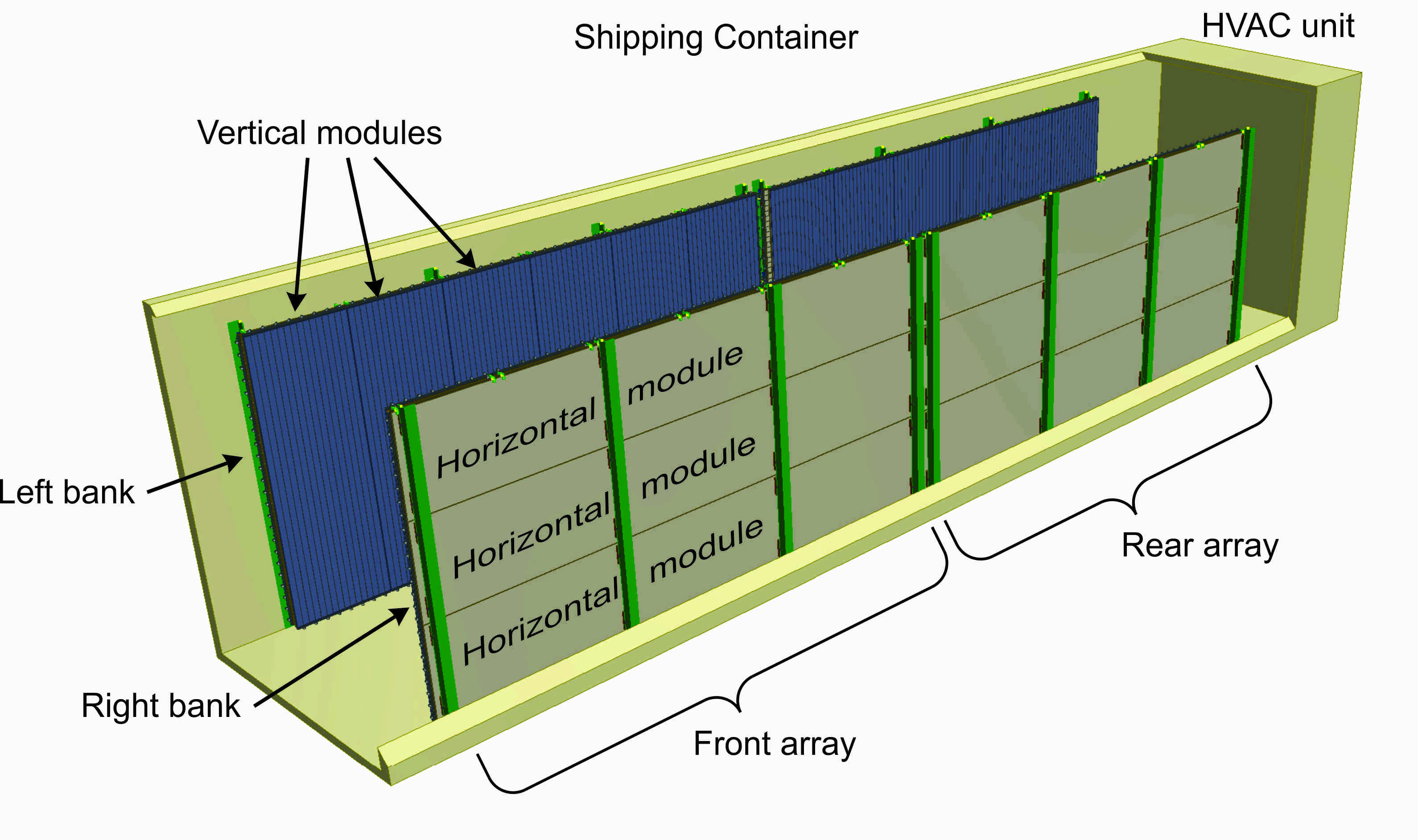}
\end{center}
\caption[Shipping container with detector]%
{Shipping container with side and top removed showing the detector layout.}
\label{fig:container}
\end{figure}

\begin{figure}[h]
\begin{center}
\includegraphics[width=0.32\textwidth]{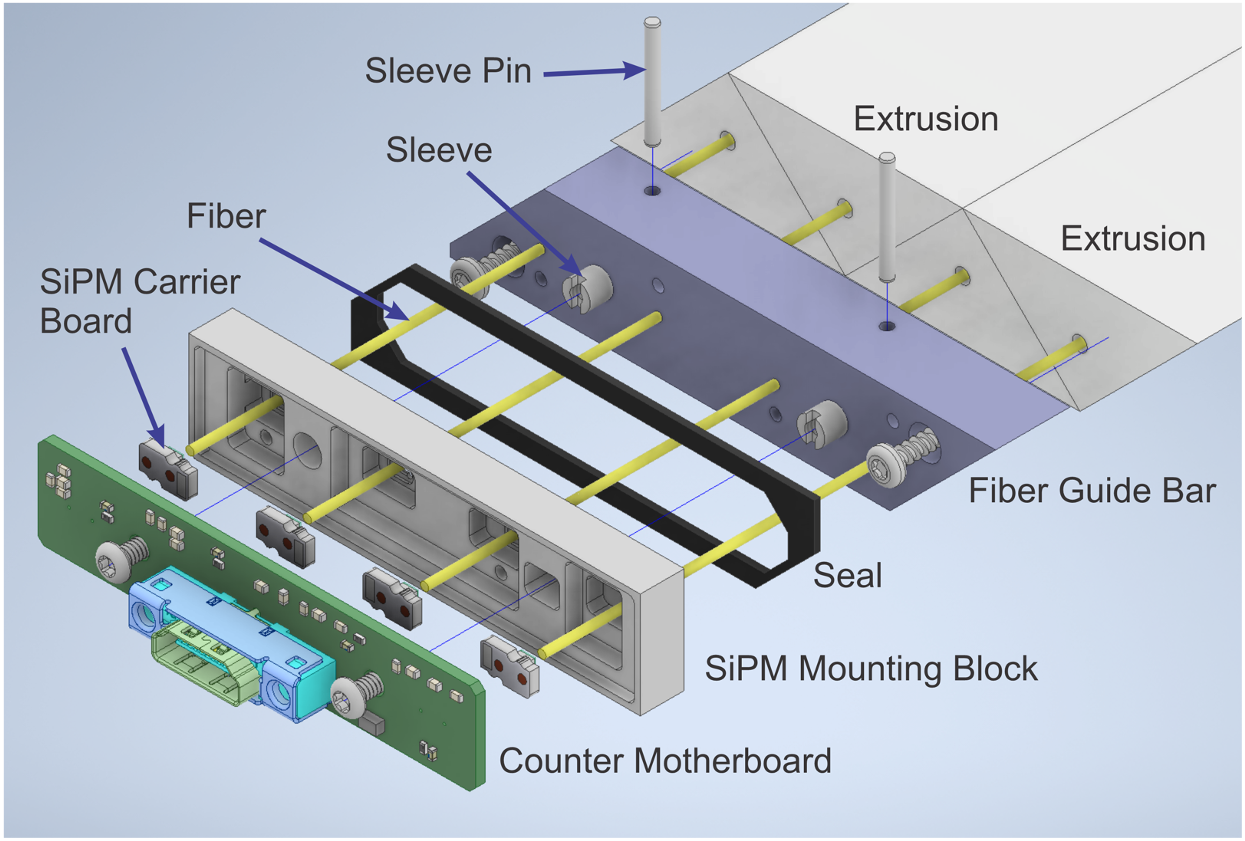}
\includegraphics[width=0.32\textwidth]{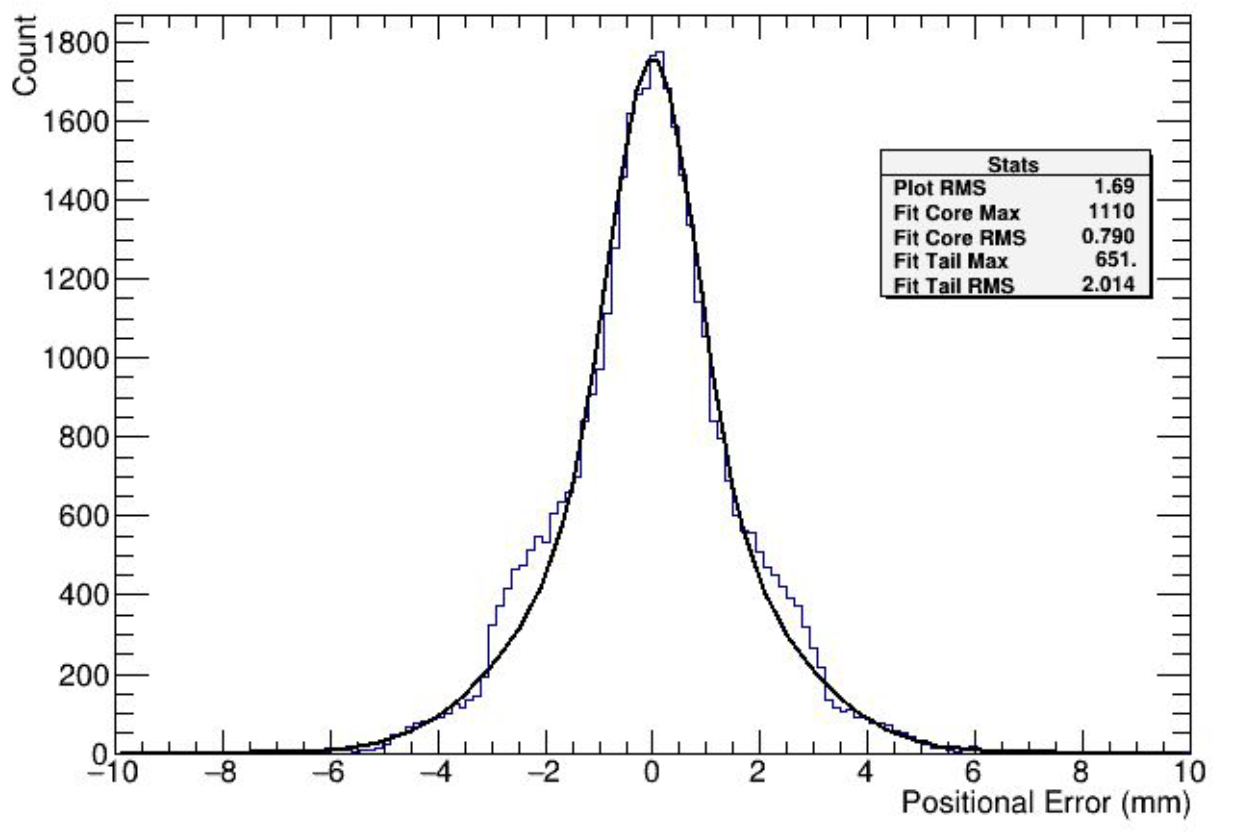}
\end{center}
\caption{Left: Exploded view of scintillator Quad-counter detector module.  Right: Simulation of point resolution using the expected light yield.}
\label{fig:dicount}
\end{figure}

The  philosophy behind our approach is to follow the concepts that were first developed for geophysical studies~\cite{Marteau:2012zv}  (plastic scintillator arrays), but make them much larger and assemble them in temperature-controlled overseas shipping containers in order to make the telescope systems very robust and easily portable.  The basic unit is 40' long, by 8' wide, by 9.5' tall.
Our baseline model for the containers has them outfitted with two scintillator banks that are constructed from horizontal and vertical modules (see Figure~\ref{fig:container}), which will provide the two views (X-Y) for each bank.  We are also exploring the use of a third bank which would allow us to determine the muon's momentum via Multiple-Coulomb Scattering (MCS). See below. Matching a hit in a horizontal module with one in a vertical module with the use of time information produces a unique point in the bank.  Connecting two points, one in the right bank and one in left bank, defines the trajectory of the cosmic-ray muon.  Collecting muon trajectories in this way allows us to build the image.  We have based our simulations on a system  of 2 telescopes with each telescope consisting of 4 containers in a 2 $\times$ 2 configuration (see Figure~\ref{fig:GModel}).  The telescopes will use $\simeq$ 1300~m$^2$ of scintillator strip detectors.  
The muon detector that we will use is based on the one developed for the upcoming $\mu$2e experiment at Fermilab~\cite{Donghia:2018duf}.  However, we have chosen to use a scintillator extrusion that is triangular in shape (isosceles with a 4 cm base and a 2 cm height) so that we get more precise point resolution.  The basic unit consists of 4 strips called a Quad-counter. See Figure~\ref{fig:dicount} (Left). When a muon passes through a counter, it will typically deposit charge in two of the triangular extrusions.  By using a charge weighting algorithm (relative amount of ionization deposited in the two strips), we can achieve a point resolution with $\simeq$ 0.8~mm core and 2~mm tail. See Figure~\ref{fig:dicount} (Right).  The side-bands shown in Figure~\ref{fig:dicount} (Right) are due to build up of the reflector in the vertices at the base of the triangle which we saw in our prototype detectors (see Section~\ref{sec:Proto}).  Although we believe that this can be fixed with further tuning of the extrusion die, in the GEANT4 simulation used to produce Figure~\ref{fig:dicount} (Right) all detector ``artifacts" where put into the model along with all physics processes.

\begin{figure}[h]
\begin{center}
\includegraphics[width=0.34\textwidth]{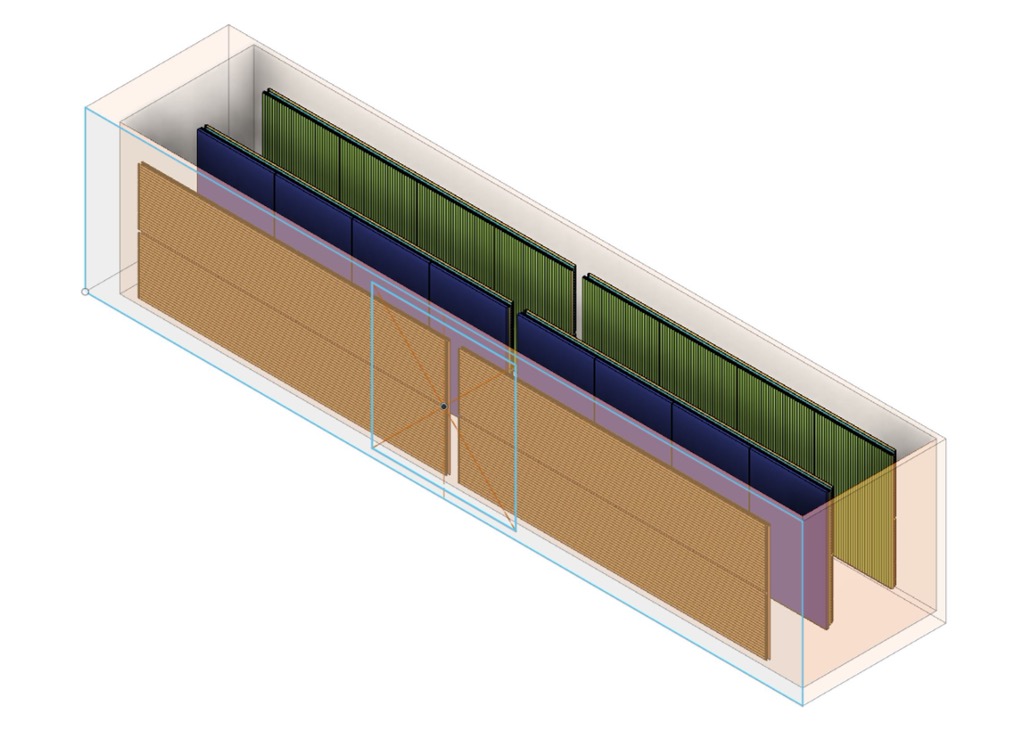}
\includegraphics[width=0.34\textwidth]{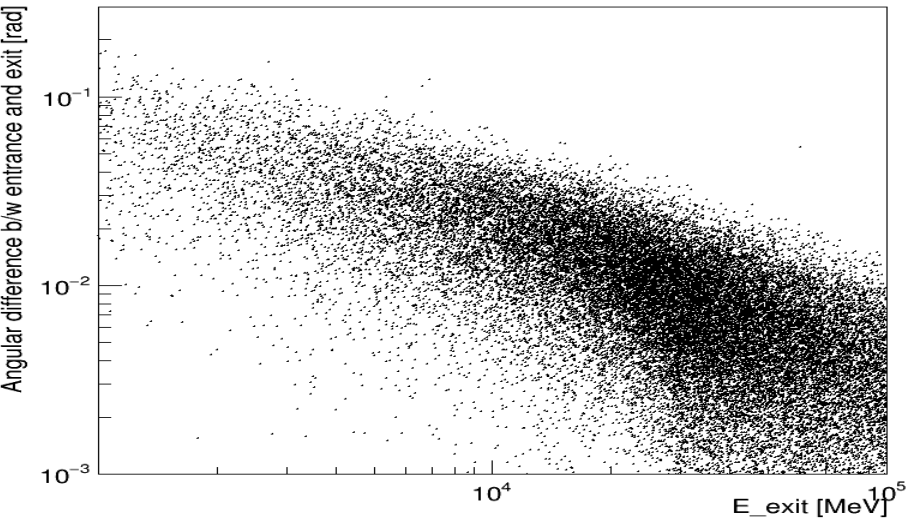}
\end{center}
\caption{Left: 3-bank configuration for the telescope.  Right: Simulation of the difference between a muon's incoming angle before it enters the pyramid and its outgoing angle after it leaves the pyramid (angle measured by our telescope).}
\label{fig:3-bank}
\end{figure}

Given the $\simeq$ 2~m separation between the banks, the pointing resolution of the telescope is $\simeq$ 1 degree.  However, for objects that are further than $\simeq$ 5~m from the surface of the pyramid, the pointing accuracy is limited by MCS.  Therefore, we have considered a telescope with 3 banks where the first two (facing pyramid) measure the trajectory.  An absorber (lead or steel) is put behind the second bank and then banks 2 and 3 measure a scattering angle with respect the incoming angle measured by banks 1 and 2.  With a known amount of absorber, the relationship between scattering angle and the muon's momentum is known.  This concept has been applied for geophysical applications~\cite{DErrico:2020xrw, olah2018high}.  A schematic for this 3-bank configuration is shown in Figure~\ref{fig:3-bank} (Left).  Figure~\ref{fig:3-bank} (Right) shows the a plot of results from one of our simulations that gives the angular differences (in radians) between a muon's entrance and exit direction as it propagates through the pyramid.  A cut to remove muons with an energy below 10 GeV would remove the muons with a large direction change which have the potential to degrade the tomographic image.  Our simulations have shown that by using the 3-bank configuration in this way we can keep 92\% of the good muons (E $>$ 10 GeV) while rejecting 79\% of the muons that fall below 10 GeV.  Verifying the efficacy of the approach will require the full tomographic reconstruction which is still a work in progress.
%
%
\subsection{Detector prototype}
\label{sec:Proto}
\begin{figure}[h]
\begin{center}
\includegraphics[width=0.5\textwidth]{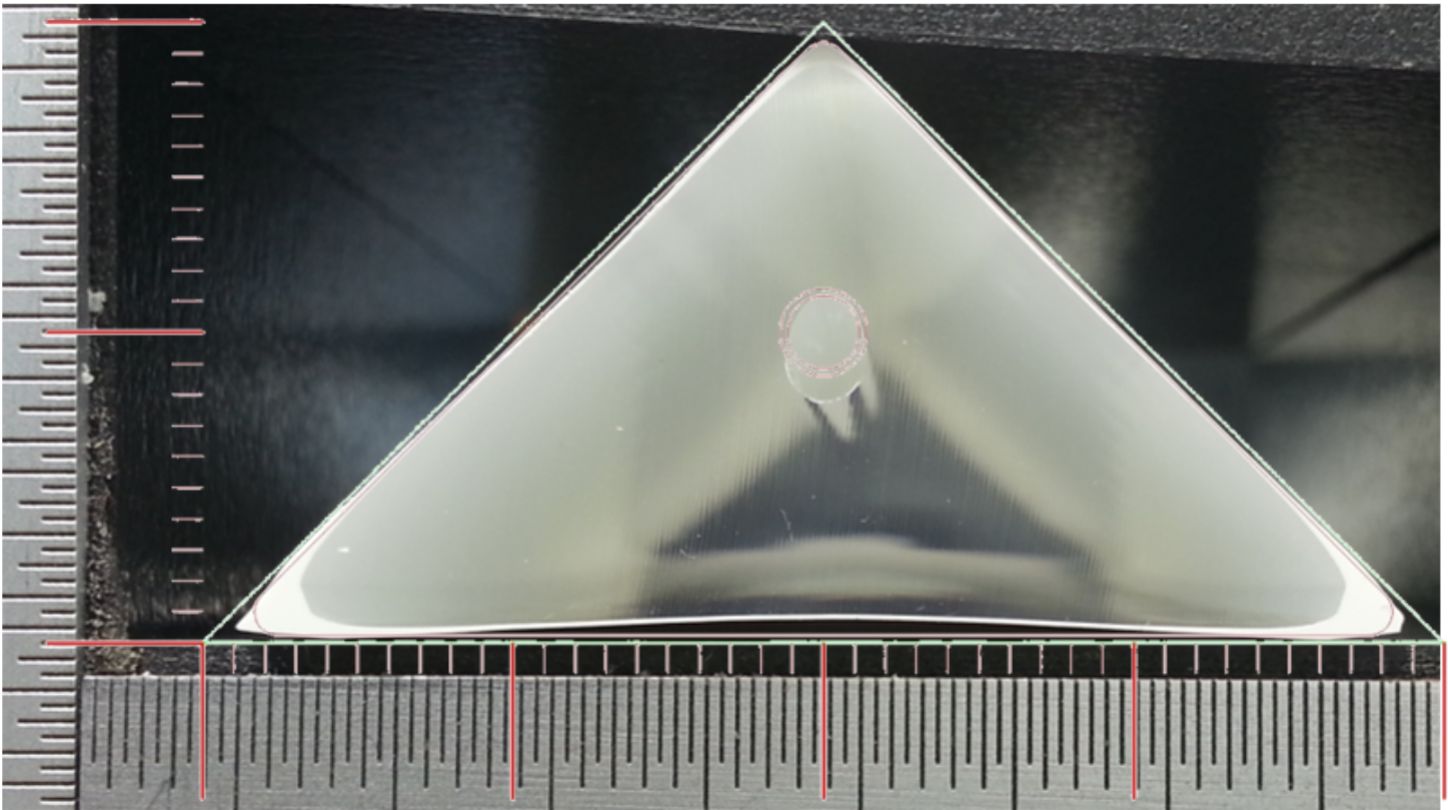}
\end{center}
\caption{Triangular extrusion from test run.}
\label{fig:Proto}
\end{figure}

Although our focus has been on the simulation work, we have begun some detector prototyping work.  New tooling for the proposed triangular extrusion has been procured and the first tuning run with this tooling has been completed.  There are some issues with the reflective coating, but overall the tooling works very well.  See Figure~\ref{fig:Proto}.  We are currently in the process of constructing the prototype telescope.  It has 4,  1~m$^2$ stations, each with an X and a Y view. This system will allow for both single-sided and double-sided muon tomography to be studied.  There are 400 strips in the system with each readout by a wavelength shifting fiber (WLS) coupled to a Hamamatsu S14283, 2 $\times$ 2~mm SiPM mounted on a custom PC board, as shown in Figure~\ref{fig:dicount} (Left).  The WLS fiber is 1.4~mm diameter Kuraray round, multi-clad fiber (non-S type) doped with Y11 at a concentration of 175 parts per million.  The baseline is to have both ends of the WLS fiber readout with a SiPM, but single-ended readout is also possible with the use of reflector module instead of the SiPM mounting block shown in Figure~\ref{fig:dicount} (Left).  We currently have all the scintillator, SiPMs and Quad-counter assembly parts in hand.
\newpage
\section{Simulation results}
\label{sec:Sim}
In order to test the capabilities of the proposed EGP telescopes, we are using the GEANT4~\cite{Ivanchenko:2003xp} toolkit and the Daya Bay cosmic-ray generator~\cite{dayabay}, which provides better agreement with data for low-energy muons than does the Gaisser model~\cite{Gaisser:2008zz}. The simulation is run on the Open Science Grid~\cite{OSG} and is split into 2 stages.  The first stage simulates a cosmic-ray muon's propagation through the pyramid up to the point where it reaches the detector.  The second stage continues the simulation where stage 1 stopped and applies track filtering and reconstruction.  Having the detector simulation separated in its own stage allows us to simulate different detector designs without re-running the time-consuming stage 1.
We are studying a number of models of the Great Pyramid that include a training model, a model including the known structure and a model including the known structure plus hypothetical features.  The training model (shown in Figure~\ref{fig:GModel}) consists of a concrete pyramid the size of the Great Pyramid and a 3-dimensional array of spherical voids with diameters between 1~m and 6~m.   For the simulation, telescopes (2 $\times$ 2 arrays) are placed fully along the 4 sides of the Great Pyramid, 36 positions in total.  This configuration yields more efficient use of CPU.  The images in grey in Figure~\ref{fig:GModel} represent moving the telescope around the pyramid to detect muons penetrating the pyramid from all angles. The results presented in this paper are for an effective 1.7-year view period for the phantom model and an effective 2.1-year view period for the Khufu model, where 2, 2 $\times 2$ arrays would periodically be moved  along the base ($\simeq$ 5 weeks in each position).
Muons in an energy range between 30 GeV and 1 TeV, polar angle between $45^\circ$ and $90^\circ$ and full azimuth angular range ($\pm 90^\circ$) with respect to the detector are used.  We simulated $\simeq 4\times10^{10}$ muons which corresponded to 43 days of viewing with the 36 telescope arrays as implemented in the simulation.  In addition to simulating $\simeq$ 2-year exposures for the models of the Great Pyramid, we also produce a ``Baseline Run" which is just a solid pyramid.
\subsection{2D reconstruction}
\label{subsec:2D}
For each muon that is detected by the telescope, the muon trajectory is calculated and projected back into the pyramid.  These ``back-projected'' tracks are then used to fill 3D histograms representing voxels of the interior of the pyramid that the track went through.  For this study, the voxel dimension was set to be 1 m$^3$.  The 3D histograms integrate over all detected muons.  The algorithm we are currently using then divides the 3D histograms for the data (model under study) by those of the Baseline.  Areas with lower density than appears in the Baseline (solid) model  will be indicated by voxels that have more entries. In the 2D plots that follow, this means a number greater than one on the color scale.  The full 3D histogram has $\simeq 8\times10^6$ voxels with $\simeq 1.6\times10^5$ hits per voxel on average.

\begin{figure}[h]
\begin{center}
\includegraphics[width=0.95\textwidth]{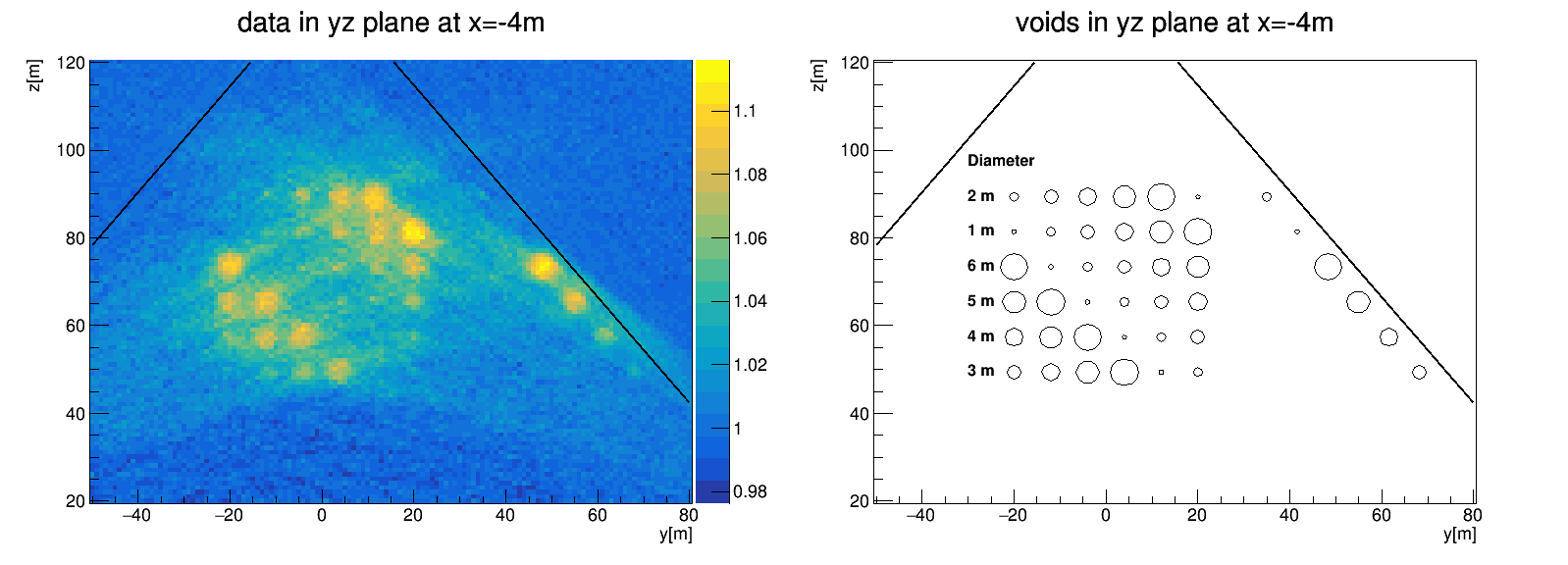}
\end{center}
\caption{Right: Phantom distribution in the model at this x location. Left: yz slice from the 3D histogram at x = 7~m.}
\label{fig:yzslice}
\end{figure}

Figure~\ref{fig:yzslice} shows data from the simulation of our training model for one yz slice through the pyramid at x = -4~m .  This is a position roughly 110~m into the pyramid.  Figure~\ref{fig:yzslice} (Right) shows the cross-section of the phantoms placed in the model in this yz plane and Figure~\ref{fig:yzslice} (Left) shows the 2D reconstruction from the muon data.  Voids as small a 3~m are visible.
As a point of reference, the large new void that the ScanPyramids team discovered after roughly 2 years of viewing was $\simeq$ 30m long $\times$ by $\simeq$~6m in diameter. This is a volume that is approximately 60 times larger than the 3~m voids that we can see in our simulation with roughly the same viewing time.

\begin{figure}[h]
\begin{center}
\includegraphics[width=0.50\textwidth]{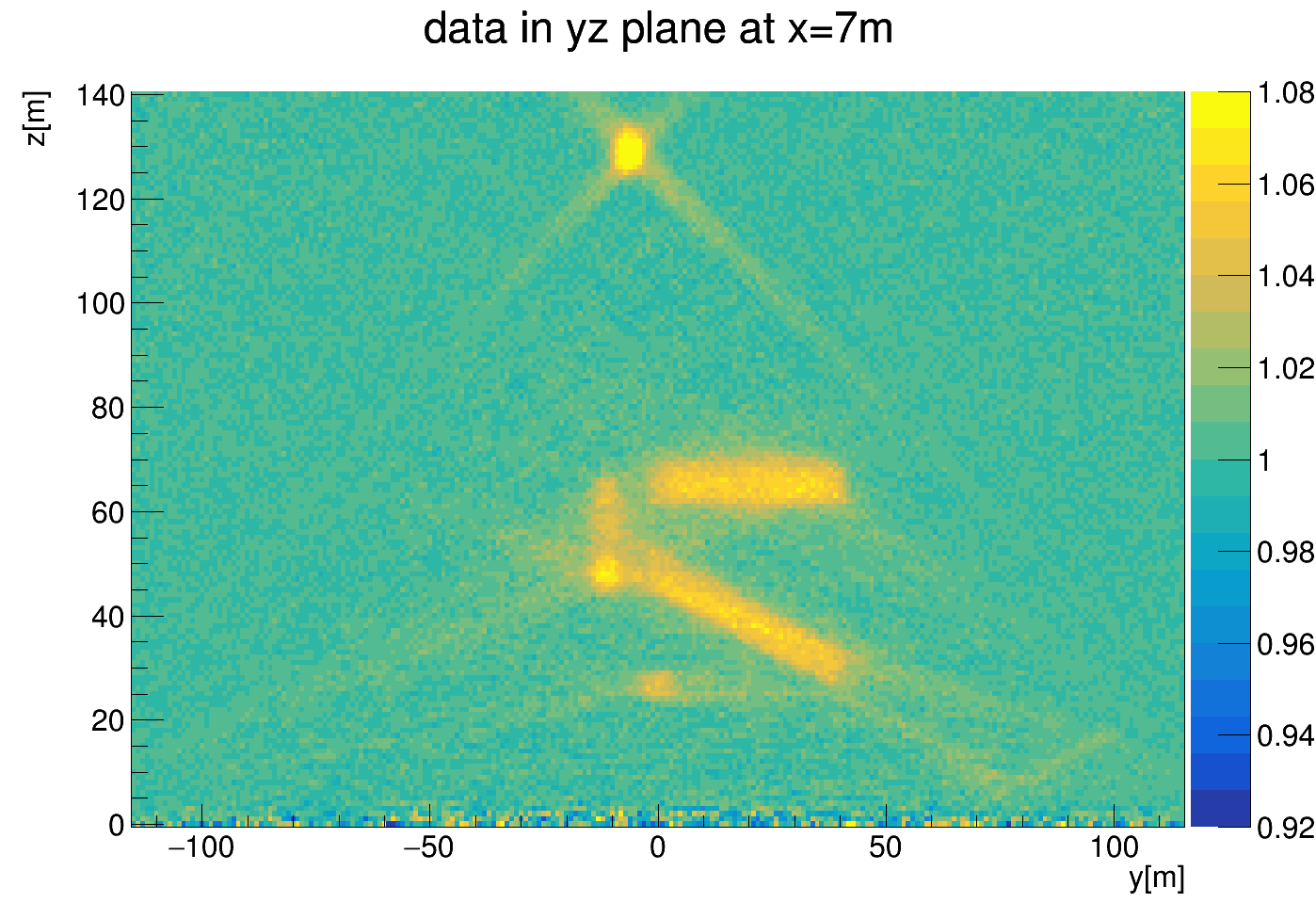}
\includegraphics[width=0.48\textwidth]{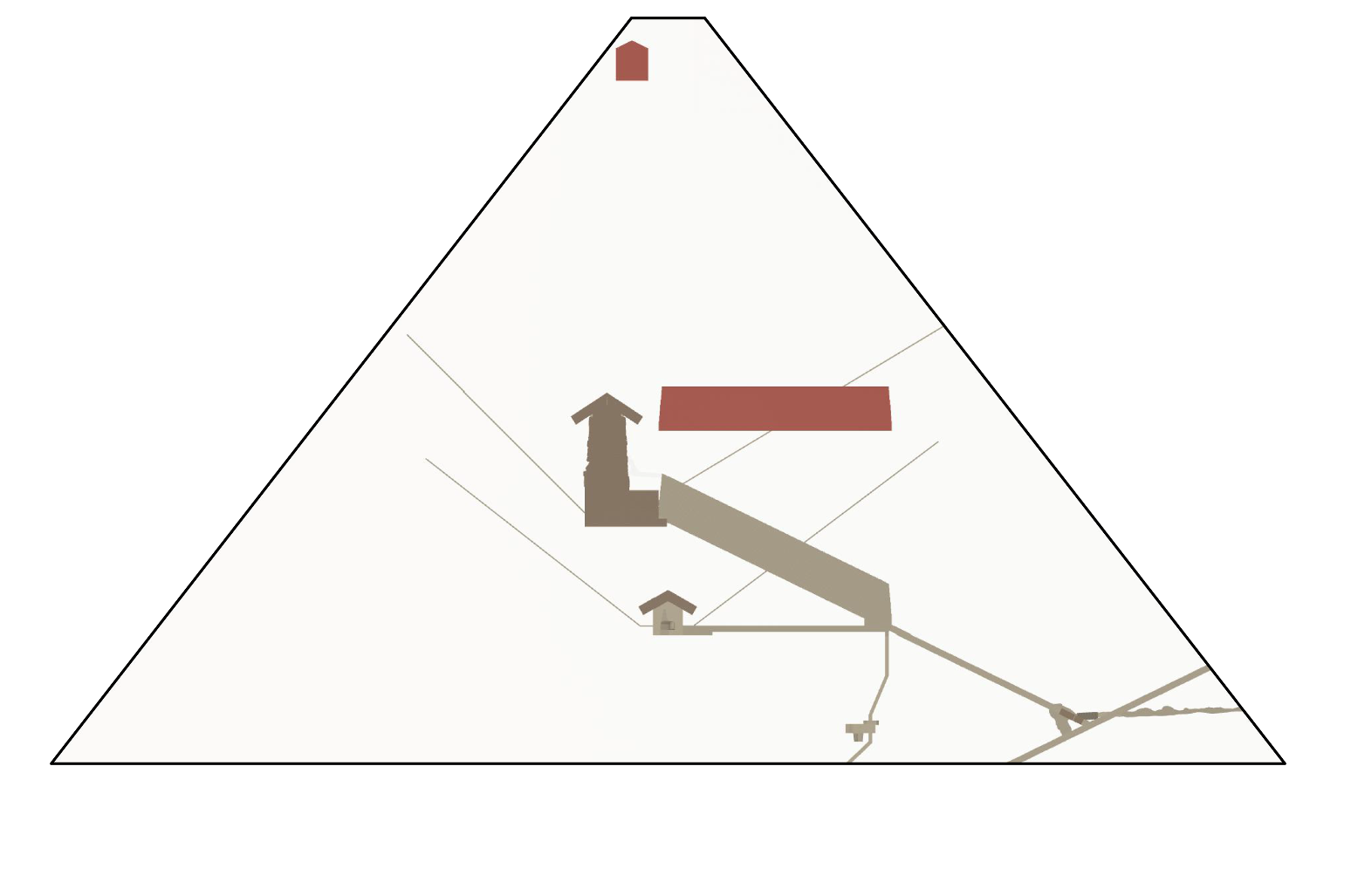}
\end{center}
\caption{Left: yz slice from the 3D histogram.   Right: Khufu model.}
\label{fig:Khufu}
\end{figure}

Figure~\ref{fig:Khufu} shows data from a simulation of one of our models of the Great Pyramid.  In addition to the well-established known structure, we have added a representation of the ``New Big Void" reported by the ScanPyramids collaboration and a second King's chamber motivated by the miro-gravity data.  Figure~\ref{fig:Khufu} (Right) shows the cross-section in the x = 7~m plane from our model and Figure~\ref{fig:Khufu} (Left) shows the 2D image reconstruction.  In the reconstruction the King's chamber, the relieving chambers above the King's chamber, the Queen's chamber and its connecting corridor, the Grand Gallery as well as the ``New Big Void" and the second King's chamber are all clearly visible.  The artifacts (rays emanating from the second King's chamber for example) are due to the back-projection and will be removed in a true 3D tomographic reconstruction (see section~\ref{subsec:3D}).  The fidelity in which we can see these structures indicates that we will be able to detect density variations at the 10-20\% level over larger volumes as indicated in micro-gravity data of the Great Pyramid.
\subsection{Tomographic reconstruction}
\label{subsec:3D}
\subsubsection*{Muon binning for tomographic reconstruction}

When the detector is parallel to the $x$ axis, we can assign a spatial
coordinate (denoted $x,z$) and a direction $\theta,\phi$ to each
detected muon. Here $\theta$ denotes the polar or zenith angle measured
from the vertical ($z$) axis and $\phi$ denotes the azimuthal angle,
represented relative to the $y$ axis. The event quadruplet
is thus $x,z,\theta,\phi$ and defines a ray in 3D space. Note that
for the case where the detectors are on the orthogonal face of the
pyramid, the event quadruplet will be $y,z,\theta,\phi$, with $\phi$
now defined relative to the $x$ axis. 

For a single container, the number of unique
such rays is given by the product of the number of 2 cm pixels in
the front and back of each detector container (note that we are assuming 2 cm pixelation of the detectors in this section with no further enhancement of the resolution using a vernier strategy or the addition of 3-station detection). The front and back
detection panels are each 960 cm $\times$ 240 cm = 480 pixels  $\times$ 120
pixels. The number of lines connecting each pixel on the back to each
pixel on the front is $(480*120)^{2}$, which is approximately $3\times10^{9}$.
A single container parked for a month is expected to detect approximately
$2\times10^{7}$ muons. There are thus substantially more rays than
detected events. We cannot use traditional tomographic reconstruction
approaches based on line integrals in this regime. Those require many
events per ray in order to invert Eqn. \ref{eq:num_muons} for the opacity line integrals.

A natural solution is to bin the raw data. The key is to reduce the
effective number of rays while preserving potential reconstructed
image quality. Geometric considerations suggest that it is critical
to maintain angular resolution and thus argue for binning mainly or
exclusively in the spatial coordinates. With 2 cm pixels separated
by 2 m, the native angular resolution is 0.0004 steradians and this
will be the target angular bin size. To sample over the entire quarter
sphere in the viewing direction of the pyramid, which spans $\pi$
steradians, will require $\pi/0.0004$=7285 directional bins. In reality
we will need fewer samples since we do not need to sweep $\theta$
through 90 degrees. We begin with 15 samples in $\theta$ from 0 to
45 degrees; the number of samples in $\phi$ varies based on $\theta$
to maintain approximately constant bin solid angle.

The spatial binning, in contrast, can be relatively coarse. We have
started with 48 $\times$ 48cm spatial bins. For containers that are 960 cm
$\times$ 240 cm this will yield 20 $\times$ 5 = 100 bins. With this
binning, one container position now samples 728,500 binned rays (assuming
the earlier calculation of 7,285 directional bins over the quarter
sphere). This is much smaller than the expected number of detected
muons. We denote the binned data $n(x_{i},z_{j},\theta_{k},\phi_{m})$
where $n$ denotes a number of muons detected along some line of sight
$L(x_{i},z_{j},\theta_{k},\phi_{m})$ through the pyramid. 

Using the methods discussed in Sec. \ref{sec:Tomo}, we can estimate the opacity
(line integral through the pyramid density map) corresponding to the
ray direction $L(x_{i},z_{j},\theta_{k},\phi_{m})$ from $n(x_{i},z_{j},\theta_{k},\phi_{m})$.
At this point, our interest is in understanding the potential image
resolution and contrast achievable from a set of ideally measured
line integrals, so we have performed our tomography simulations and
reconstructions in density and opacity space, without passing through
the nonlinear transform mapping opacity to numbers of detected muons.
The latter will be important in the future when we wish to study the
effects of Poisson counting noise and muon spectrum miscalibration
errors. 

\subsubsection{Tomographic image reconstruction}

The goal is to use this set of opacity samples to estimate the 3D
density map of the pyramid., which we discretize into voxels. We choose
a discretization of 250 X 250 X 160 voxels of size 0.96m (deliberately
chosen to be a multiple of the spatial binning interval). This is
$\sim1\times10^{7}$ voxels spanning 240m X 240 m X 153.6 m, which
is large enough to contain the pyramid. Each binned ray can be traced
through the voxel grid and the intersection length of the ray with
each voxel calculated. To do so we use Siddon's method \cite{siddon1985fast}.
This determines which voxels contribute to each measurement and with
what weight. 

Gathering the opacity estimates $\varrho(x_{i},z_{j},\theta_{k},\phi_{m})$
into a vector $\mathbf{g}$ and the voxel values into a vector $\mathbf{f}$, we can then link them in a matrix equation,
\begin{equation} \label{system_equation}
\mathbf{g}=H\mathbf{f}.
\end{equation}

With, for example, 10 container positions along each of two sides of the pyramid,
we will have a total of $\sim1.5\times10^{7}$ opacity
measurements, which exceeds the number of unknown voxel values. $H$
is a matrix of the intersections lengths with the voxels. Note that
$H$ is an enormous matrix. It is $\sim10^{7}\times10^{7}=10^{14}$
elements and can't be stored as a whole, but it has a lot of sparsity,
structure, and redundancy that significantly reduces computational
and storage requirements. 

For image reconstruction itself, we are investigating the use of the
SIRT algorithm, which was found by Hartling et al.~\cite{hartling2021comparison} to be the most
promising in the context of muon tomography of a nuclear reactor core.
The SIRT update is given by

\begin{equation} \label{SIRT}
\mathbf{f}^{(k+1)}=\mathbf{f}^{(k)}+\frac{\lambda_{k}}{\sum_{i}h_{ij}}\sum_{i}\frac{\mathbf{g}_{i}-h_{i}\mathbf{g}_{i}^{(k)}}{\sum_{j}h_{ij}}h_{i}^{T},
\end{equation}
where $\lambda_{k}$is a relaxation parameter, the $h_{ij}$ are the
elements of $H$, and $\mathbf{h}_{i}$ denotes the vector corresponding to
row $i$ of $H$.

\begin{figure}[h]
\begin{center}
\includegraphics[width=0.25\textwidth]{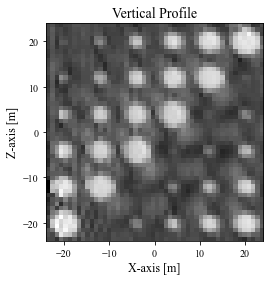}
\includegraphics[width=0.25\textwidth]{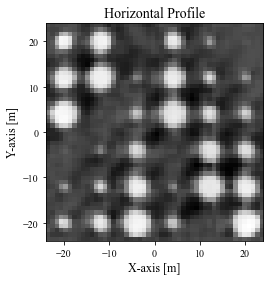}
\end{center}
\caption{Left: Vertical slice from the 3D SIRT reconstruction of a phantom comprising spheres ranging in size from 1 m to 6 m. Voxel size is 96 cm and 200 iterations were employed. }
\label{fig:Spheres}
\end{figure}

We performed a preliminary 3D SIRT reconstruction of an arrangement of spheres ranging in size from 1 m to 6 m, arrayed in the same pattern as those shown in the central portion of Figure~\ref{fig:yzslice}. We modeled a single stack of two containers being shifted uniformly to 25 position along each of two faces of the pyramid. Each container is 9.6x2.4 m with 480x120  detector elements of size 0.02x0.02m. After reparameterizing in spatial and angular coordinates, the spatial coordinates were binned 24x24 fold creating 0.48x0.48 meter spatial bins. The angular binning  created 120 polar angles (theta) and 960 azimuthal angles (phi) at each translation. The reconstructed grid was voxelixed into (250,250,160) voxels with a voxel size of .96x.96X.96 m. The algorithm was run for 200 iterations. The results, shown in vertical and horizontal cuts in Figure~\ref{fig:Spheres}, allow clear visualization of spheres 2 m and larger. These reconstructions also show reduced streak artifacts relative to the histogramming approach shown in Figure~\ref{fig:Spheres}, which is essentially an unfiltered backprojection reconstruction. Note that these tomography simulations contained no noise, model mismatch, or other systematic errors, and serve to assess the spatial resolution achievable by the detection geometry and reconstruction approach under idealized conditions. 

\section{Outlook and conclusions}

\label{sec:Outlook}
The Exploring the Great Pyramid Mission takes a different approach to imaging large structures with cosmic-ray muons.  The use of very large muon telescopes placed outside the structure, in our case the Great Pyramid of Khufu on the Giza plateau, can produce much higher resolution images due to the large number of detected muons.  In addition, by moving the telescopes around the base of the pyramid, true tomographic image reconstruction can be performed for the first time.  The detector technology employed in the telescopes is well established and prototyping of specific components has already begun.

Initial simulation results for the EGP Mission provide convincing evidence that the concept can provide powerful new insight into structural details of the Great Pyramid of Khufu.  We have shown that all of the known structure (exclusive of the air shafts) can been seen in our 2D reconstruction in addition to the ``New Big Void" and a hypothetical second King chamber put into our model.  Initial results with the application of full tomographic reconstruction show the expected improvement over 2D reconstruction, but the reconstruction is still under development.  We expect that full implementation and optimization of the tomographic reconstruction will improve the imaging capability of the EGP telescopes further.
\section*{Acknowledgements}
This document was prepared by the Exploring the Great Pyramid Mission using funds provided by the University of Chicago's Big Ideas Generator and resources of the Fermi National Accelerator Laboratory (Fermilab), a U.S. Department of Energy, Office of Science, HEP User Facility. Fermilab is managed by Fermi Research Alliance, LLC (FRA), acting under Contract No. DE-AC02-07CH11359.

\bibliographystyle{unsrt}
\bibliography{EGP}

\end{document}